\documentclass[aps, prl, reprint, superscriptaddress, amsmath, amssymb, floatfix, longbibliography]{revtex4-2}

\usepackage{physics}
\usepackage[normalem]{ulem}
\usepackage{graphicx}
\graphicspath{ {pdffig/}, {supfig/}, {images/}, {figs/} }
\usepackage{dcolumn}
\usepackage{bm}
\usepackage{hyperref}
\usepackage{siunitx} 
\usepackage{ulem}    
\usepackage{pifont}
\usepackage{xcolor}
\hypersetup{
	colorlinks,
	linkcolor={blue!90!black},
	citecolor={blue!90!black},
	urlcolor={blue!90!black}  
}
\usepackage[caption=false]{subfig}
\makeatletter
\newcommand{\colorcaption}[2][]{%
  \begingroup%
  \renewcommand{\@caption@fignum@sep}{ (color online). }%
  \caption[#1]{#2}%
  \endgroup%
}
\makeatother

\def\Rb{$^{87}$Rb}
\def\Na{$^{23}$Na}
\def\K{$^{40}$K}
\def\NaRb{$^{23}$Na$^{87}$Rb }

\begin{document}


\title{Seconds-scale coherence on nuclear spin transitions of ultracold polar molecules in 3D optical lattices}

\author{Junyu Lin}
\affiliation{Department of Physics, The Chinese University of Hong Kong, Shatin, Hong Kong, China}
\author{Junyu He}
\affiliation{Department of Physics, The Chinese University of Hong Kong, Shatin, Hong Kong, China}
\author{Mucan Jin}
\affiliation{Department of Physics, The Chinese University of Hong Kong, Shatin, Hong Kong, China}
\author{Guanghua Chen}
\affiliation{Department of Physics, The Chinese University of Hong Kong, Shatin, Hong Kong, China}
\author{Dajun Wang}
\email{djwang@cuhk.edu.hk}
\affiliation{Department of Physics, The Chinese University of Hong Kong, Shatin, Hong Kong, China}
\affiliation{The Chinese University of Hong Kong Shenzhen Research Institute, Shenzhen, China}

\date{\today}

\begin{abstract}

Ultracold polar molecules (UPMs) are emerging as a novel and powerful platform for fundamental applications in quantum science. Here, we report characterization of the coherence between nuclear spin levels of ultracold ground-state sodium-rubidium molecules loaded into a 3D optical lattice with a nearly photon scattering limited trapping lifetime of 9(1) seconds. After identifying and compensating the main sources of decoherence, we achieve a maximum nuclear spin coherence time of $T_2^* = 3.3(6)$~s with two-photon Ramsey spectroscopy. Furthermore, based on the understanding of the main factor limiting the coherence of the two-photon Rabi transition, we obtain a Rabi lineshape with linewidth below 0.8 Hz. The simultaneous realization of long lifetime and coherence time, and ultra-high spectroscopic resolution in our system unveils the great potentials of UPMs in quantum simulation, computation, and metrology.


\end{abstract}

\maketitle


Following the rapid progress in creating and controlling ultracold polar molecules (UPMs) in recent years~\cite{KRbstirap,KRb2008,RbCs201411,RbCs201412,NaK201505,NaRb201605,NaLi201710,NaK201801,NaK2020,DirectSrF,CaFbelowDoppler,HighDensityCaF,3DMOTYO,subdopplerYO}, many of the assets essential for fulfilling their long-held promises in a broad range of applications are finally starting to be explored. The strong and long-range dipole-dipole interaction (DDI) between UPMs placed in DC electric fields has been exploited to modify the chemical reaction~\cite{Ni2010,valtolina2020dipolar} and complex formation rates~\cite{GuoPRX}. The vibrational level has been used as a switch to control the two-body chemical reactivity~\cite{ye2018collisions}. The rotational degree of freedom has been investigated in even more details. Electric field induced F\"{o}rster resonance between UPMs in their first excited rotational level are used to suppress chemical reactions for obtaining trapped samples with long lifetimes~\cite{KRbResonanceShield} and for reaching quantum degeneracy via direct evaporative cooling~\cite{li2021tuning}. Effects of the resonant DDI naturally occurred between UPMs in rotational levels with opposite parities have been observed with~\cite{NaKResonantCol} and without the microwave coupling~\cite{NaRbResonantCol}. 

Quantum information processing is one of the first proposed applications for UPMs. While the earlier proposals rely on the DDI induced by external DC electric field for quantum gate operation~\cite{QuantumComputationwithPolar,Schemesforrobustcom}, currently the resonant DDI has become the more popular choice~\cite{ni2018dipolar,Sawant2020,Hughes2020}. In this paradigm, two adjacent rotational levels, which can be manipulated with high fidelity by microwave signals readily available in ultracold systems, serve as the spin up and spin down states of the molecular qubit. Significant efforts have been made to understand and improve the rotational coherence time for both samples in optical dipole traps~\cite{KRbMagicAngle,NaKMagictrap,NaRbRotationalCoherence} and in 3D optical lattices~\cite{spinexchangeKRb}. In the latter case, decoherence caused by long-range dipolar spin exchange has been observed. For single molecules trapped in the optical tweezer, rotational coherence time of 93 ms, which is long enough for many two qubits gate operations, was measured very recently~\cite{CaFRotationalCoherence}.

The nuclear spin hyperfine structures of UPMs are another valuable resource for quantum information processing. Pairs of nuclear spin states in the same rotational state, which are insensitive to external fields and free from decoherence due to DDI, can be used as storage qubits. In two recent experiments, nuclear spin coherence times of 0.7 s for $^{23}\rm{Na}^{40}\rm{K}$ molecules~\cite{NaKHyperfineCoherence} and 5.6 s for $^{87}\rm{Rb}^{133}\rm{Cs}$ molecules~\cite{RbCsHyperfineCoherence} have been reported. However, these experiments were performed with bulk samples which are subject to significant two-body losses due to two-molecule complex formation~\cite{KRbComplexes,LossofUltracoldRbCs,NaKBrightAndDark,NaKNaRbLightInduced} even though both of them are chemically stable. For the next step forward, it is thus necessary to shield the molecules from each other.

In this work, we suppress the two-body loss by directly loading UPMs into deep 3D optical lattice potentials. For the \Na\Rb\, molecules used here, a trapping lifetime of 9 s is observed in a very low filling lattice. With this long-lived sample, long coherence time is measured using two-photon microwave Ramsey spectroscopy for several pairs of nuclear spin states. In the optimized case with minimized differential Zeeman and AC Stark shifts, coherence time over 3 s is observed. In this case, we also discover that the damping of the two-photon Rabi oscillation is limited by the differential one-photon detunings across the sample and slows down for smaller Rabi frequencies. This has allowed us to drive long Rabi pulses with Fourier transform limited linewidths below 0.8 Hz which should be enough to resolve few and many-body effects caused by DDI over distances of several lattice sites.

\begin{figure}[t]
    \centering
    \includegraphics[width=0.9\linewidth]{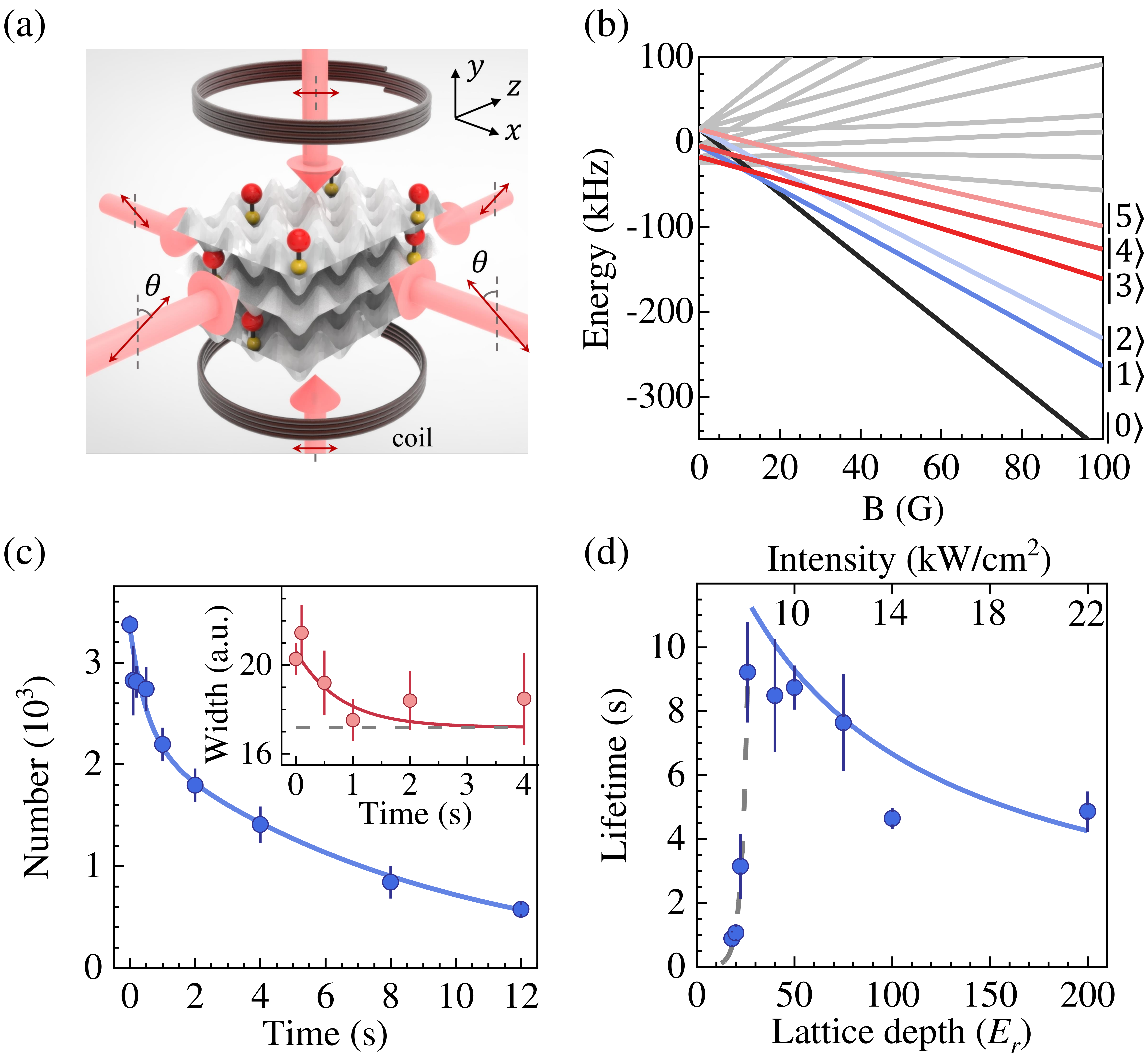}
\caption{
    Long-lived ground-state \NaRb molecules in 3D optical lattices. 
    (a) The optical lattice is formed by three pairs of linearly polarized laser beams. The polarization angle $\theta$ of the horizontal lattice beam can be adjusted from 0$^{\circ}$ to 180$^{\circ}$. For the vertical lattice beams $\theta$ is fixed at 90$^{\circ}$. 
    (b) The $J = 0$ level has 16 nuclear spin Zeeman states. The 6 states labeled from $\ket{0}$ to $\ket{5}$ can be populated individually by STIRAP. 
    (c) Evolution of the molecule number in a sparsely filled lattice with a trap depth of 50$E_{\rm{r}}$ per beam. The solid curve is the second-order exponential fitting which shows a fast initial decay and a longer lived tail. 
    Inset: Evolution of the lattice band occupation measured with the band mapping technique. The gray dashed line shows the simulated width for molecules occupying the ground band. The red solid curve is an exponential fitting for eye guiding. 
    (d) Lifetime versus lattice trap depth. The gray dashed curve is from the exponential fitting for lattice depth smaller than 26$E_{\rm{r}}$, while the blue solid curve is the calculated off-resonant light scattering limited lifetime. Error bars are one standard error.
    } 
\label{fig1}
\end{figure}

The experiment starts from about 7000 ultracold ground-state \Na\Rb\, molecules prepared in a single nuclear spin state of the $J = 0$ rotational level by population transferring of Feshbach molecules via the stimulated Raman adiabatic passage (STIRAP)~\cite{NaRb201605}. By adjusting the Raman laser parameters, six different nuclear spin states labeled in Fig.~\ref{fig1}(b) with $\ket{0}$ to $\ket{5}$~\cite{Note1} in order of increasing energy at 100 G can be reached. 
The sample is held in an optical dipole trap formed by two 1064.2 nm laser beams crossing each other at $90^\circ$ in the horizontal plane and has a typical initial temperature of 210~nK and a mean density of $10^{10}$~cm$^{-3}$. To load the molecules into the 3D lattice, the retroreflections of the two horizontal beams, as well as the vertical lattice [Fig.~\ref{fig1}(a)] are ramped on within 30 ms after the STIRAP. As shown in Fig.~\ref{fig1}(c), due to two-body losses during the loading process, only about 50\% of the 7000 molecules are loaded into the optical lattice even with this relatively short loading time. With the low initial density and the loss, the lattice can only be filled sparsely. The estimated mean lattice filling factor is only 0.1\%. The probability of having more than one molecules on the same site is thus very small.

In the 50$E_{r}$ per beam lattice used in taking the data in Fig.~\ref{fig1}(c), the oscillation frequencies are about 23 kHz. Here $E_{r}$ is the photon recoil energy of \Na\Rb\, at 1064.2 nm. For the typical initial sample temperature, most of the molecules occupy the ground band of the lattice after the loading. However, the molecules in the higher band, which have faster tunneling rates, can tunnel into sites already occupied and lead to two-body losses~\cite{KRb3Dlattice}. Due to the tight lattice confinement, the effective density of two molecules in the same site is on the order of $10^{13}$~cm$^{-3}$. With a typical loss rate constant of $10^{-10}$~cm$^3$s$^{-1}$~\cite{ye2018collisions}, the two-body loss rate is on the order of $10^3$ s$^{-1}$ which is much faster than the tunneling rate, and the observed loss rate is thus determined by the tunneling rate of the higher band molecules. This picture is consistent with the observed fast loss in the initial 1.5 s which reduces the number of molecules by about 30\% [Fig.~\ref{fig1}(c)]. As shown by the band occupation measured with the band mapping method [inset of Fig.~\ref{fig1}(c)], this initial fast loss serves as a purification process after which nearly all remaining molecules are in the ground band~\citep{KRb3Dlattice}. With the tunneling now negligible, the loss of the remaining molecules is much slower and follows a one-body curve with a 1/e lifetime of $\tau = 8.7(7)$~s.

To understand the limiting factors to the lifetime, we further investigate its dependence on the lattice depth. Figure~\ref{fig1}(d) shows the lifetime of the slow decay part for lattice depths from $18E_{r}$ to $200E_{r}$ per beam. From $18E_{r}$ to $26E_{r}$, the lifetime increases rapidly with increasing lattice depth which we attribute to the suppression of the tunneling rate. For even larger lattice depths, the lifetime starts to decrease. The blue solid curve is the lifetime calculated with the theoretical imaginary polarizability $\rm{Im}(\alpha) = 1.85\times 10^{-5}~\rm{a.u.}$ at 1064.2 nm~\cite{theoryPolari}. We note that different from atoms, here each photon scattering event will lead to the loss of one molecule. The nice agreement between theory and experiment supports that similar to the case in \K\Rb\,\cite{KRb3Dlattice} off-resonance scattering from the trapping light is also the main limiting factor of the lifetime here.

%

\begin{figure}[t]
    \centering
    \includegraphics[width=1.0\linewidth]{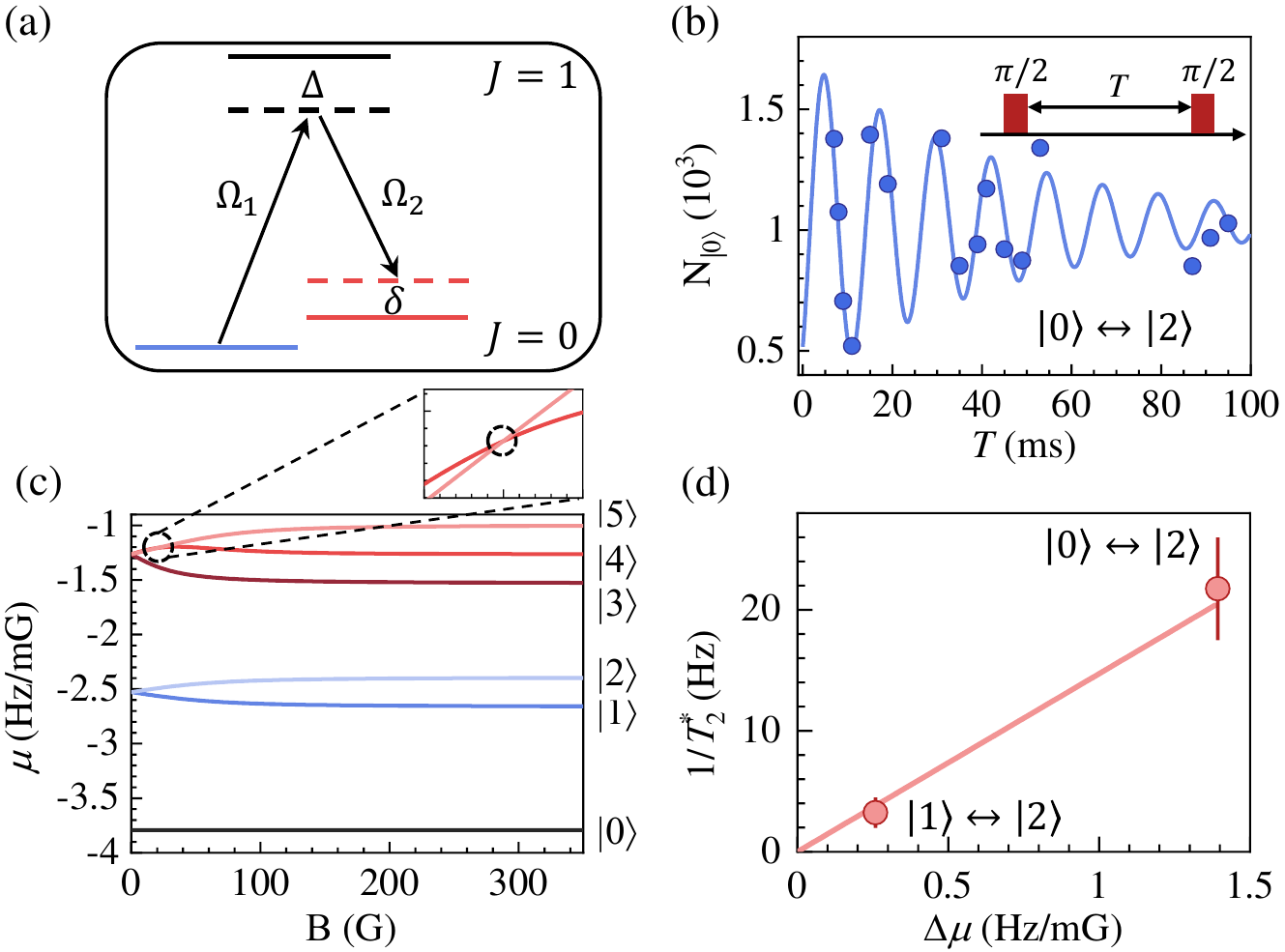}
    \caption{Measuring the nuclear spin coherence time $T_2^*$ with two-photon Ramsay spectroscopy.
    (a) Two nuclear spin states in $J = 0$ are coupled together by a two-photon microwave (MW) transition with one-photon detuning $\Delta$ and two-photon detuning $\delta$. $\Omega_1$ and $\Omega_2$ are the one-photon Rabi frequencies.      
    (b) The Ramsey precession between $\ket{0}$ and $\ket{2}$ at 335 G observed by measuring $N_{\ket{0}}$. The solid curve is from the fitting to Eq.~\ref{eq:fitting} for extracting the $T_2^*$.
    (c) The magnetic dipole moment $\mu$ of the 6 nuclear spin states for magnetic field from 0 to 350 G. Zoomed in section: at 20 G, $\ket{4}$ and $\ket{5}$ have the same $\mu$.  
    (d) The decoherence rate 1/$T_2^*$ has a clear dependence on $\Delta \mu$. The slope of the linear fitting forced through the zero point is determined by the magnetic field noise. 
    }
    \label{fig2}
\end{figure}

With the long-lived sample at hand, now we investigate the coherence between nuclear spin states in $J = 0$ with Ramsey spectroscopy. As shown in Fig.~\ref{fig2}(a), these nuclear spin states can be coupled together by two-photon microwave transitions. Figure~\ref{fig2}(b) is the Ramsey precession between $\ket{0}$ and $\ket{2}$ at 335 G. For this measurement, the intermediate state for the two-photon transition is a hyperfine level in $J = 1$ with mixed nuclear spin characters which allow it to couple to both $\ket{0}$ and $\ket{2}$~\cite{NaRbinternalstate}. Starting from a pure $\ket{0}$ sample after the STIRAP, a coherent superposition state is created by a $\pi/2$-pulse with a small two-photon detuning $\delta$. After a variable free precession $T$, a second $\pi/2$-pulse is applied and the number of molecules in $\ket{0}$ ($N_{\ket{0}}$) is measured. To extract the nuclear spin coherence time $T_2^*$, we fit the Ramsey precession to
\begin{equation}
    N_{\ket{0}} = \frac{N_0}{2} e^{-{T}/{\tau}}[ 1 + A_0 e^{-{T}/{T_2^*}} \cos(\delta t + \phi_0)], 
\label{eq:fitting}
\end{equation}
with the initial molecule number $N_0$, the oscillation amplitude $A_0$, the initial phase $\phi_0$, and $T_2^*$ as the fitting parameters. We emphasize that all the coherence measurements are performed in the long-lived part of the molecule number evolution. For the set of data in Fig.~\ref{fig2}(b), the independently measured trap lifetime $\tau$ is 7(1) s. From the fitting, a coherence time $T_2^* = 46(9)~\rm{ms}$ is obtained.

One of the main limitations to $T_2^*$ is the differential Zeeman shift between the nuclear spin states. As shown in Fig.~\ref{fig2}(c), the small magnetic dipole moment $\mu$ can still cause energy shifts on the order of 1 Hz per mG. More importantly, the magnetic dipole moment difference $\Delta\mu$ between two nuclear spin states typically is not zero. For states $\ket{0}$ and $\ket{2}$ at 335 G, $\Delta\mu$ is 1.4 Hz/mG. Thus, magnetic field noise on the 10 mG level will limit $T_2^*$ to below 100 ms. To verify this understanding, we measure the coherence time between $\ket{1}$ and $\ket{2}$ in exactly the same magnetic field and optical lattice configurations. The only difference is that $\Delta\mu$ between $\ket{1}$ and $\ket{2}$ has a smaller value of 0.26 Hz/mG. Indeed, a much longer $T_2^*$ of 309(120)~ms is observed. With the two $T_2^*$, a fitting to $1/T_2^* = \delta B \Delta\mu$ forced through the origin [Fig.~\ref{fig2}(d)] gives a magnetic field noise $\delta B$ of 15(1) mG. This is consistent with the combined contributions of the magnetic field noise and gradient measured from other methods.


\begin{figure}[t]
\centering
  \includegraphics[width=0.95\linewidth]{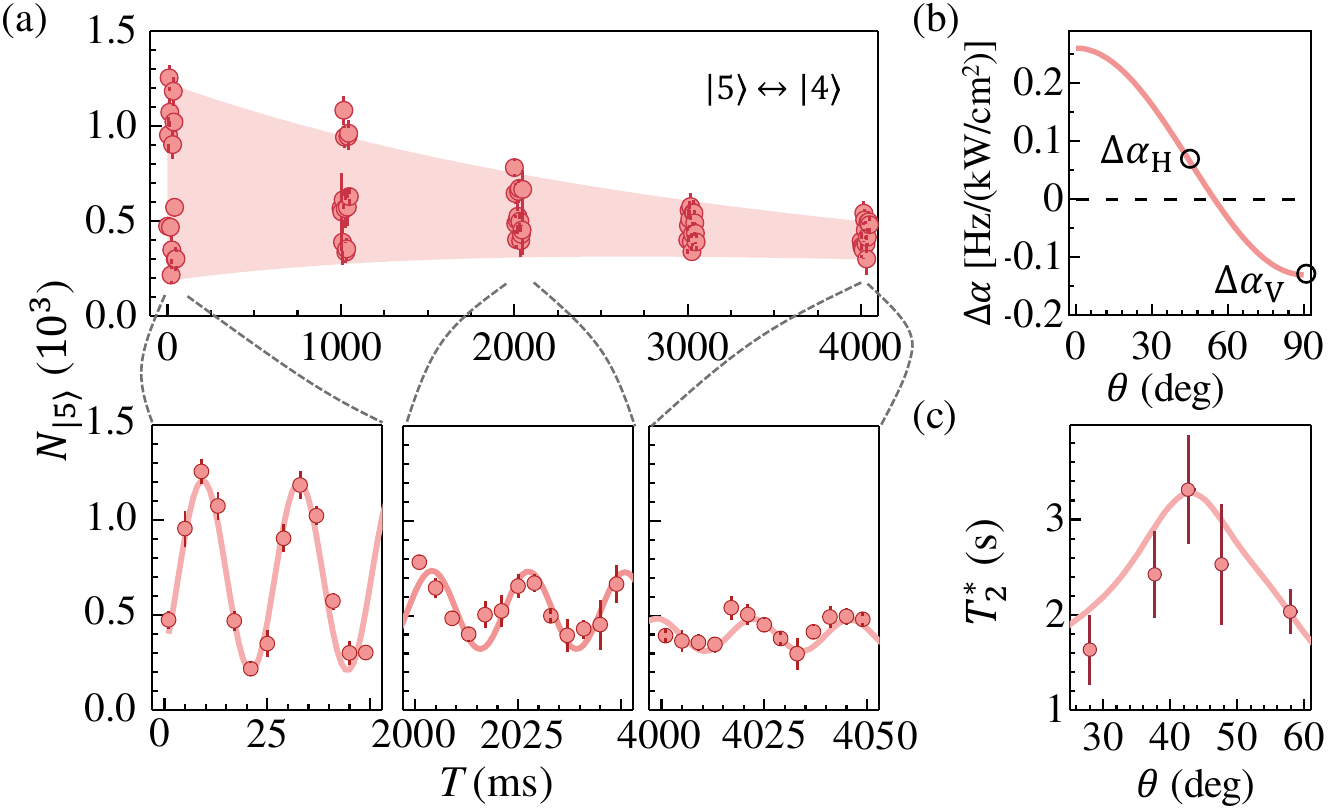}
\caption{
    Seconds-long coherence time between nuclear spin states $\ket{4}$ and $\ket{5}$ at 20 G.
    (a) Ramsey precession obtained by measuring the number of molecules in $\ket{5}$. The solid curve is from fitting to Eq.~\ref{eq:fitting} with an extracted $T_2^*$ of 3.3(6) s. The polarization angles of the horizontal lattice beams are $\theta = 43^{\circ}$. Bottom: magnified sections of the data set above. 
    (b) The difference between the anisotropic polarizabilities of $\ket{4}$ and $\ket{5}$ as function of $\theta$. The two black circles labeled as $\alpha_{\rm{H}}$ and $\alpha_{\rm{V}}$ indicate $\theta$ for the vertical and horizontal lattice beams used in (a) for compensating the differential AC stark shifts between $\ket{4}$ and $\ket{5}$.
    (c) $T_2^*$ measured for several polarization angles $\theta$ of the horizontal lattice beams. The solid curve shows the result of a Monte Carlo simulation. All error bars represent standard error.
    }
\label{fig3}
\end{figure}

A special case is the pair of states $\ket{4}$ and $\ket{5}$ which at 20 G have the same $\mu$ and thus a zero $\Delta\mu$, as illustrated in Fig.~\ref{fig2}(c) and the zoomed in section. Near this point, the differential Zeeman shift between $\ket{4}$ and $\ket{5}$ induced by the magnetic field noise becomes negligible. Figure~\ref{fig3}(a) presents the Ramsey precession in this special situation obtained by measuring the number of molecules in $\ket{5}$ ($N_{\ket{5}}$). In the magnified sections, clear coherent oscillations can still be observed after $T = 4$~s. The $T_2^*$ is measured to be 3.3(6) s, which is 70 times longer than that between $\ket{0}$ and $\ket{2}$ at 335 G. However, to obtain this very long coherence time, we have found that it is necessary to use a specific lattice light polarization combination. As shown in Fig.~\ref{fig3}(c), with the polarization angle of the vertical lattice beams fixed at 90$^\circ$, the longest coherence time only occurs when the polarization angle of the horizontal lattice beams are tuned to 43$^\circ$. This indicates that there are differential AC Stark shifts between the nuclear spin states and, contrary to our typical understanding~\cite{NaRbRotationalCoherence}, the AC polarizabilities of $J = 0$ nuclear spin states are also anisotropic.

The origin of this polarizability anisotropy in $J = 0$ is the nuclear electric quadrupole interaction which can induce coupling between $J = 0$ to $J = 2$ in presence of light fields~\cite{NaKHyperfineCoherence,RbCsHyperfineCoherence}. This leads to a small anisotropic polarizability term on top of the background isotropic polarizability $\alpha_0$. In the current parameters, $\Delta \alpha$, the difference between the anisotropic terms of two nuclear spin states is less than $2\times 10^{-5}$ of $\alpha_0$. This causes a Hz level differential AC Stark shift across the whole sample. Thus, its constrain on $T_2^*$ will only become observable when $T_2^*$ reaches 1 s level. Figure~\ref{fig3}(b) shows the light polarization dependence for $\Delta \alpha$ between $\ket{4}$ and $\ket{5}$ at 20 G calculated using the polarizabilities and hyperfine coupling parameters obtained previously~\cite{NaRbRotationalCoherence}. Ideally, $\Delta \alpha$ can be tuned to zero by setting the polarization angle to $54.7^{\circ}$. However, the polarization angle of the vertical lattice beams in our experiment can only be $90^\circ$, and their contributions to $ \Delta \alpha$ is thus fixed at $-0.13~\rm{Hz/(kW/cm^2)}$ [$\Delta \alpha_V$ in Fig.~\ref{fig3}(b)]. To minimize the differential AC Start shifts, we are forced to set the polarization angles of the two pairs of horizontal lattice beams to $\theta  = 43^\circ$ where their contributions to $\Delta\alpha$ [$\Delta \alpha_H$ in Fig.~\ref{fig3}(b)] can largely counterbalance those from the vertical beams. The solid curve in Fig.~\ref{fig3}(c) is from a Monte Carlo simulation of $T_2^*$ with the Gaussian beam profiles of the lattice beams and the gravitational sag taken into account which agrees with the measured $T_2^*$ for several $\theta$ very well.

\begin{figure}[t]
    \centering
    \includegraphics[width=0.9\linewidth]{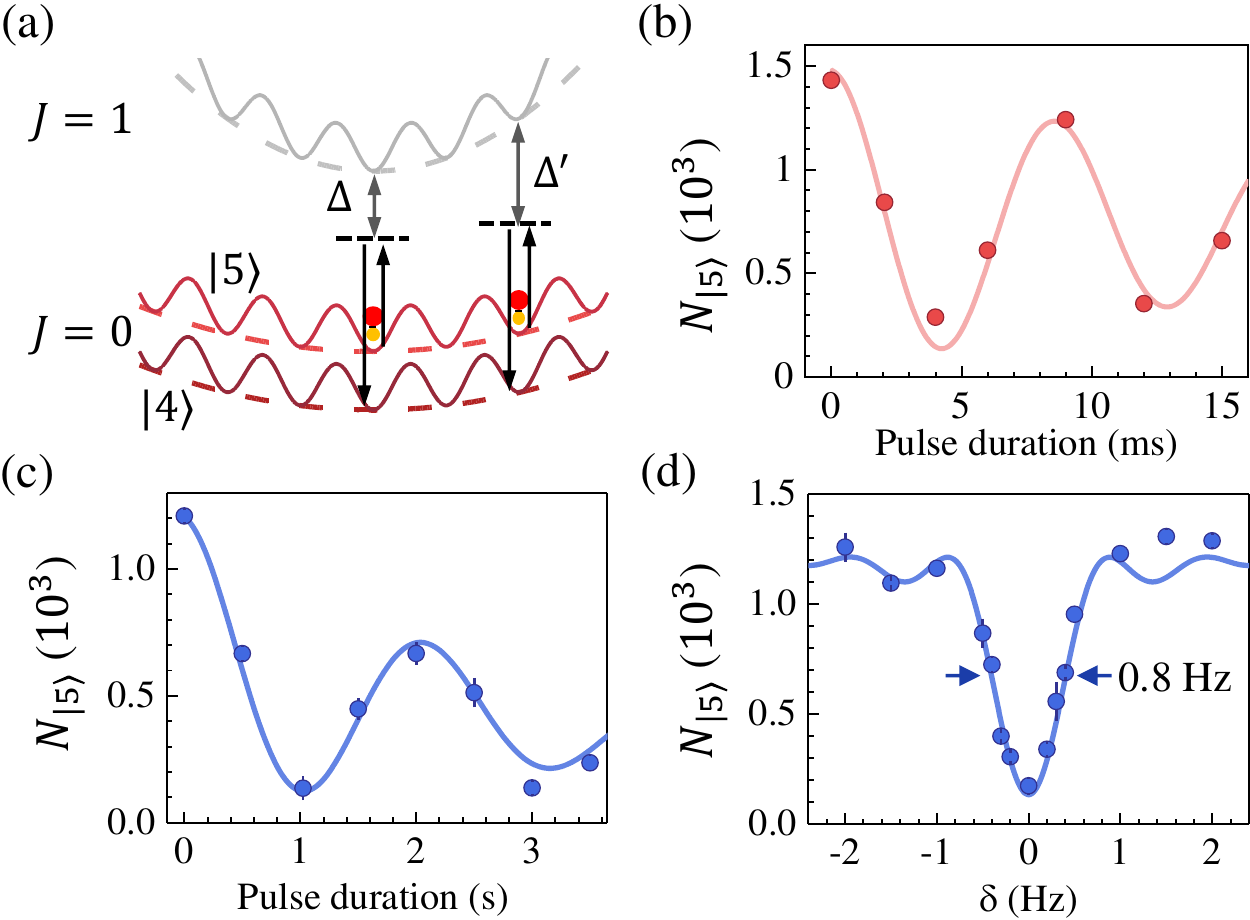}
    \caption{Two-photon Rabi spectroscopy with sub-Hz linewidth.
    (a) Schematic lattice potentials. While the differential AC Stark shift between $\ket{4}$ and $\ket{5}$ are largely compensated, that between $J = 0$ and $J =1$ is still relatively large. Consequently, the one-photon detunings ($\Delta$ and $\Delta'$) are spatially dependent and the two-photon Rabi frequency is also not uniform. 
    (b) and (c) Damped two-photon Rabi oscillations between $\ket{4}$ to $\ket{5}$ at 20 G obtained with two-photon Rabi frequencies of about $2\pi\times 115$ Hz and  $2\pi\times 0.5$ Hz, respectively.
    (d) A two-photon Rabi lineshape between $\ket{4}$ to $\ket{5}$ obtained with a one-second pulse. An ultra-narrow linewidth of 0.8 Hz is achieved. 
    }
    \label{fig4}
\end{figure}

The seconds-long coherence time observed here is already enough to make the nuclear spin states of \NaRb a possible candidate for long-time quantum information storage~\cite{Schemesforrobustcom,ni2018dipolar}, as well as for many other applications. Here, we demonstrate their potential in quantum metrology with the sub-Hz linewidth two-photon Rabi spectroscopy. As illustrated in Fig.~\ref{fig4}(a), due to the spatially-dependent differential AC Stark shift between $J = 0$ and $J = 1$, the one-photon detuings $\Delta$ also have different values across the sample. This is the primary factor limiting the coherence time in the rotational transition driven by a one-photon microwave~\cite{NaRbRotationalCoherence}. For the transition between nuclear spin stats here, the two-photon Rabi frequency $\Omega = \Omega_1 \Omega_2/2\Delta$ also becomes spatially dependent which results in obvious damping of the two-photon Rabi oscillations [Fig.~\ref{fig4}(b) and (c)]. However, as $\Omega$ is proportional to the product of the one-photon Rabi frequencies, its variation across the sample and thus the damping rate will become smaller at reduced $\Omega_1$ and  $\Omega_2$. Indeed, comparing the Rabi oscillations in Fig.~\ref{fig4}(b) and (c) (note the different time scales in their horizontal axes), the damping rate in Fig.~\ref{fig4}(c), which is obtained with $\Omega = 2\pi \times 0.5$~Hz, is much slower than that in Fig.~\ref{fig4}(b) taken with $\Omega = 2\pi \times 115$~Hz. As shown in Fig.~\ref{fig4}(d), this allows us to apply a one-second interrogation pulse which results in a Rabi lineshape with Fourier transform limited linewidth of 0.8 Hz and center frequency resolution of $\pm 0.01$ Hz.   

The 3.3(6) s coherence time is still far from the fundamental limit of the coherence between $\ket{4}$ and $\ket{5}$ at 20 G. 
In the current polarization combination, the differential AC Stark shift is minimized but is not eliminated. This issue can be further mitigated by replacing the vertical lattice with an optical accordion~\cite{Li2008,Williams2008} or a tightly focused single beam. In both methods, the laser beams propagate mainly in the horizontal direction so that its polarization angle, as well as those for the horizontal lattice beams, can be adjusted to 54.7$^{\circ}$ to make $\Delta \alpha$ zero. The trapping lifetime could also be increased by changing the lattice light to a longer wavelength. At 1550 nm, for instance, the photon-scattering limited lifetime will be more than 30 s at the same trap depth. With all these improvements, reaching $T_2^*$ longer than 10 s should be feasible.

To conclude, we have created an ultracold sample of ground-state \Na\Rb\, molecules confined in a 3D optical lattice with long lifetime and long nuclear spin coherence time. Depending on the choice of nuclear spin state pairs and magnetic fields, the coherence time can span from tens of milliseconds to several seconds. Coupled with the strong DDI for fast gate operation and the long trap lifetime, this makes \Na\Rb\, in optical lattice a nice choice for quantum information processing.

In the current study, very low lattice filling factors are used so that the DDI does not cause observable effects on the nuclear spin coherence. However, this could be dramatically improved, for example, by creating a double Mott insulator of \Na\, and \Rb\, first and then performing the molecule association~\cite{RbCsLattice}. For \Na\Rb\, in optical lattices with modest filling factors of 20\% to 30\%, the two-photon nuclear spin transition will become a very sensitive probe for few and many-body dipolar physics~\cite{QuantumMagnetism1,QuantumMagnetism2,micheli2006toolbox}. For instance, in external DC electric fields, the DDI between UPMs across several lattice sites can still bring observable changes to the Rabi lineshape~\cite{PolarMoleculesWithE,RydbergBlockadeEfield,RydbergExcitationEfield}. This gives us a spectroscopic method for studying the Ising model with long-range interactions~\cite{IsingLongrange1,IsingLongrange2}. It is also possible to place some amount of the population in a carefully chosen nuclear spin state of $J = 1$ which interact strongly only with one of the nuclear spin states of $J = 0$ via the resonant DDI~\cite{RydbergDDI,RydbergDDIinLattice}. The two-photon nuclear spin transition can then be used to probe the effect of the spin exchange in the XY model without the limitation due to the short rotational coherence time~\cite{spinexchangeKRb}. Varying the $J = 1$ population, this may also become an interesting impurity problem with long-range interactions~\cite{impurityRb}.

\begin{acknowledgments}
This work was supported by Hong Kong Research Grants Council (RGC) General Research Fund (Grants No. 14301119, and No. 14301818) and the Collaborative Research Fund (Grants No. C6005-17G, and No. C6009-20GF).

\end{acknowledgments}

\appendix


%

\end{document}